\documentclass{article}

\usepackage[colorlinks,urlcolor=blue,linkcolor=blue,citecolor=blue]{hyperref}
\usepackage{color}
\usepackage[frozencache,cachedir=.]{minted}
\usepackage{graphicx}
\usepackage{siunitx}
\usepackage{hyperref}
\usepackage[margin=2cm]{geometry}
\usepackage{authblk}

\begin{document}

\title{JulianA.jl - A Julia package for radiotherapy}

\newbox{\myorcidaffilbox}
\sbox{\myorcidaffilbox}{\large\includegraphics[height=1.7ex]{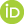}}
\newcommand{\orcidaffil}[1]{%
  \href{https://orcid.org/#1}{\usebox{\myorcidaffilbox}\,#1}}

\author[1, 2]{Renato Bellotti \orcidaffil{0000-0002-2702-0437}}
\author[1, 2]{Antony J.\ Lomax}
\author[3]{Andreas Adelmann \orcidaffil{0000-0002-7230-7007}}
\author[2,*]{Jan Hrbacek}

\affil[1]{ETH Zürich}
\affil[2]{Center for Proton Therapy, Paul Scherrer Institut}
\affil[3]{Accelerator Modelling and Advanced Simulations group, Paul Scherrer Institut}

\affil[*]{Corresponding author: jan.hrbacek@psi.ch, Forschungsstrasse 111, 5232 Villigen PSI, Switzerland }

\maketitle

\begin{abstract}
The importance of computers is continually increasing in radiotherapy. Efficient algorithms, implementations and the ability to leverage advancements in computer science are crucial to improve cancer care even further and deliver the best treatment to each patient. Yet, the software landscape for radiotherapy is fragmented into proprietary systems that do not share a common interface. Further, the radiotherapy community does not have access to the vast possibilities offered by modern programming languages and their ecosystem of libraries yet.

We present JulianA.jl, a novel Julia package for radiotherapy. It aims to provide a modular and flexible foundation for the development and efficient implementation of algorithms and workflows for radiotherapy researchers and clinicians. JulianA.jl can be interfaced with any scriptable treatment planning system, be it commercial, open source or in-house developed. This article highlights our design choices and showcases the package's simplicity and powerful automatic treatment planning capabilities.
\end{abstract}

\maketitle
\section{Introduction}
The goal of radiotherapy is to provide individualized cancer care for each patient by killing tumor cells with radiation. The early advances in the field were driven by progress in particle accelerator technology, but much of the advancements in the last two decades were enabled by increasingly powerful computer hardware and algorithms. The rise of machine learning promises to accentuate the importance of computer science in radiotherapy even further.

%Despite all this progress, much of the power of modern software ecosystems and computing hardware is still not available to medical physics researchers and practitioners. The publicly available software underlying radiotherapy research has not developed significantly over the last decade and modern software engineering practices are not widely adopted.

The crucial piece of software for any radiotherapy treatment is the treatment planning system (TPS), which is usually a commercial certified medical product. A TPS is used to display the threedimensional computed tomography (CT) images of the patient, draw the contours of the tumor and healthy organs that should be spared, perform the treatment planning (i.\ e.\ sculpt the threedimensional dose distribution in a semi-automatic way) and perform the entire decision making throughout the treatment process.

These systems are built primarily for domain experts like radiation oncologists and dosimetrists, but they offer only limited customisation via scripting through application programming interfaces (APIs). There are open source TPS like matRad \cite{matrad} and CERR \cite{cerr}, but these systems are built on MATLAB software stacks, which are not the best choice for scientific computing anymore and require expensive licenses to unlock all the features. Apart from matRad and CERR, there is OpenTPS implemented in Python~\cite{wuyckens2023}. All of these open source projects aim to provide a TPS with reduced functionality for research purposes, but there are legal obstacles to use them for treatment due to lack of certification, which hinders translation of research to clinic.

This report introduces JulianA.jl\footnote{The name is a portmaneau between Julia and FIonA, the name of our in-house TPS.} \cite{bellotti2023, code}, a Julia \cite{julia} package for radiotherapy. Instead of competing with commercial TPS, it aims to abstract their functionality and provide a uniform interface across vendors. The goal of JulianA.jl is to be as easy to use as possible for medical physics researchers to quickly develop and improve algorithms and build fully automatic workflows. The TPS abstraction layer enables workflows that include treatment planning and dose calculation across TPS vendors, which has not been possible so far.

\clearpage
\section{Challenges and Opportunities}

The complex and interdisciplinary nature of radiotherapy and its large computational demands render it a challenging and exciting application domain for computer and computational science. The following challenges and opportunities have motivated us to develop JulianA.jl.

Many heterogeneous and complex systems need to work together and conform to the highest quality standards. The complexity of systems is illustrated by the typical radiotherapy workflow: First, a CT image of the patient is acquired, sent to a medical database server and transferred to a contouring software to delineate the tumor and healthy organs. Afterwards, the contours are imported back into the database and transferred to the TPS. There, the treatment planning and quality assurance (QA) are performed in a semi-automatic way. The validated treatment plan is sent to the delivery control system. All of these systems need to interoperate in a most reliable way to enable efficient treatment workflows.

Further, a broad range of algorithms is needed, from ray casting and Monte Carlo methods for dose calculation, to uncertainty quantification and multi-objective optimisation and multi-objective decision making and even simulation models for investigating cellular response to radiation. All of these algorithms need to be implemented efficiently using modern hardware accelerators such as graphics processing units (GPUs) due to the large problem sizes encountered when working with medical images. For example, a typical CT is of size $512 \times 512 \times 100$~pixels.

The demand for automation in clinical workflows has motivated all major TPS vendors to include scripting APIs into their products. However, the APIs are not compatible with each other and differ in provided functionality and typically can only access relatively superficial functions of the underlying TPS. The incompatibility between system enforces many manual steps in clinical and research workflows, introducing inefficiencies and potential for mistakes. It even prohibits applications that rely on fully automatic workflows, such as machine learning validation pipelines.
Further, changing vendor means rewriting scripts and workflows and reimplementing algorithms from scratch. This lowers productivity, is not sustainable and fosters vendor lock-in.

Finally, the field of radiotherapy is inherently multi-disciplinary and dominated by radiation oncologists, medical physicists and dosimetrists with extensive domain knowledge, but little to no education in computer science. The importance of software is mostly neglected. In fact, software is often considered a byproduct of research, with little in the way of formal development techniques being used. Therefore, the code is written for a very specific use case, then discarded or stored on a shared drive until the next scientist or student needs to modify it. Many institutes lack a strategy for developing, maintaining, collecting or sharing of code, so new researchers need to start writing their code from scratch. Many institutes decide against sharing their scripts, either because they do not see the value of sharing or because there is limited knowledge about software engineering practices.

Modern software engineering practices can alleviate or resolve all of these issues. Therefore, radiotherapy represents an attractive application domain for computer and computational scientists. This article presents the Julia package JulianA.jl, which is an attempt to bring the power of modern software and hardware to the realm of radiotherapy. While JulianA is currently supporting only proton therapy TPS, the software is general enough to allow interfaces to the more conventional x-ray therapy TPS in the form of plugins.

\section{Scope and Goals}
JulianA.jl is designed to resolve the previously mentioned shortcomings and enable domain experts such as medical physicists to leverage the power of modern software and hardware to deliver the best possible treatment to each patient. We recognise that software engineering is only a small (yet crucial) part of clinical operation and research. Therefore, installing, learning and extending JulianA.jl must be as simple as possible.

Domain experts should be productive and try out new ideas easily, without learning how to access databases or write GUI code. For this reason, and to simplify deployment and improve versatility and code reuse, we decided to implement JulianA.jl as a Julia package, with a focus on the scripting interface rather than a GUI. In our experience, software packages tend to be more modular and therefore easier to maintain, extend and improve than GUI-based applications. JulianA.jl does not aim to provide a one-size-fits-all tool, but to provide building blocks from which specialised applications can be built.

Many institutes do not distinguish between prototype and production code due to resource constraints. Therefore, JulianA.jl aims to be as fast as possible out of the box. Capabilities to leverage modern hardware such as GPUs is mandatory, but must be accessible to domain scientists without learning a new chain of compilers, debugging tools, runtime environments or similar.

Any radiotherapy tool must ultimately be targeted towards clinical application. Therefore, JulianA.jl strives for seamless integration into existing TPS systems rather than mimicking a basic version of their functionality like existing open source TPS. Our package relies on the same certified TPS that are also used for patient treatment. This encompasses mainly physical modelling, such as dose calculation engines. Such an integration allows researchers and clinicians to build algorithms on top of the TPS. The results can be exported to standard file formats and imported back into the research or clinical TPS, where they can be validated using the usual certified QA workflows. We believe this approach unites the best of both worlds - vendor-agnostic implementation of algorithms, reliability of tried and tested TPS and well-established QA workflows. Further, it elegantly relieves domain experts from undergoing certification procedures for their code because the results are integrated into certified software.

In order to maximise productivity, JulianA.jl is implemented in the Julia programming language~\cite{julia} with a vast ecosystem of scientific packages in order to provide easy and direct access to the latest advances in computer and computational science. The ecosystem is accessible by an easy-to-use package manager that handles dependency management and allows to document and reproduce entire environments. The programming language does not require deep computer science knowledge, but allows for easy debugging and profiling, i.\ e.\ it is a high-level language.

Summarising, JulianA.jl's design renders it a well-structured foundation for advanced research endeavours and flexible clinical workflows. Thanks to the clear structure of the code and the well-defined TPS integration, a clear, clean and simple transition path from research to clinical application is given.

\section{Implementation in Julia}
We argue that Julia is the natural and obvious choice of programming language for radiotherapy. It is designed for scientific applications, provides interactive development through a powerful read-eval-print loop (REPL) and/or Jupyter notebooks, is fast due to the underlying powerful just-in-time compiler and the code is both easy to write and read thanks to a simple and elegant syntax. Julia was designed with scientific productivity in mind, resulting in a wide array of tools for debugging and profiling that integrate well into the language itself. Libraries are extensible and flexible thanks to the multiple dispatch paradigm \cite{julia}. Access to hardware accelerators is easy and unified for all vendors thanks to \texttt{KernelAbstractions.jl} \cite{kernelabstractions.jl} and the underlying GPU-vendor-specific libraries. Julia includes a powerful package management system that automatically documents environments such that the package can be instantiated on other machines with a single command. Finally, there is a vast ecosystem of libraries for scientific computing. This ecosystem extends beyond the Julia language itself because Python and C libraries can be called from Julia and Julia functions can be called from those libraries.

\section{The TPS Interface}
\begin{figure*}
    \centering
    \includegraphics[width=\textwidth]{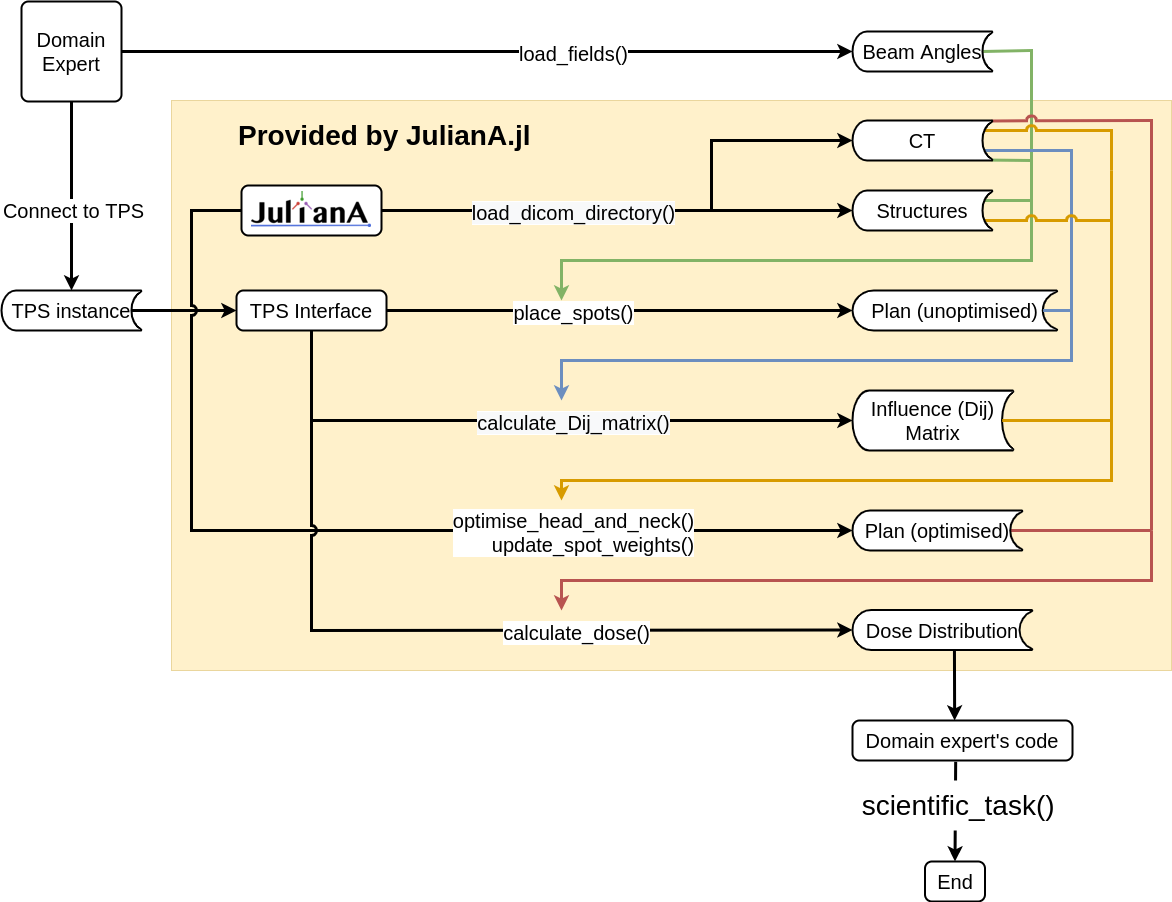}
    \caption{Dataflow diagram of JulianA.jl code from a domain expert's perspective. Squares represent whether JulianA.jl, the TPS interface or the domain expert (i.\ e.\ user) provide the functionality. Bookmark-shaped boxes indicate data structures. The function \texttt{scientific\_task()} is a placeholder for additional steps required to obtain a scientific result.}
    \label{fig:architecture}
\end{figure*}

The TPS interface decouples algorithms, e.\ g.\ for automatic treatment planning or machine learning, from the underlying physics engine. This architecture increases modularity, renders the code more easily testable, boosts code reuse and allows researchers to benchmark and compare their algorithms for various TPS. While existing open source TPS cannot be used in a clinical environment due to lack of certification, JulianA.jl benefits directly from TPS physics engines that are certified for treatment. We believe this unique feature will accelerate the translation of improved algorithms to the clinic. Fig.~\ref{fig:architecture} displays how user code connects to JulianA.jl and the TPS interface. The domain expert(i.\ e.\ the user of JulianA.jl) only has to define irradiation beam angles that are used to postition the irradiation device, call JulianA.jl code and finally receive a dose distribution that can be further processed to solve a domain expert's scientific question. Therefore, all of the physics-specific code is encapsulated by JulianA.jl and all of the user code can be run on any treatment planning system for which a JulianA.jl integration exists. Due to the interoperability of the Julia language, user code can be written not only in Julia, but also in Python and C++.

\begin{figure}
    \centering
    \includegraphics[width=0.5\textwidth]{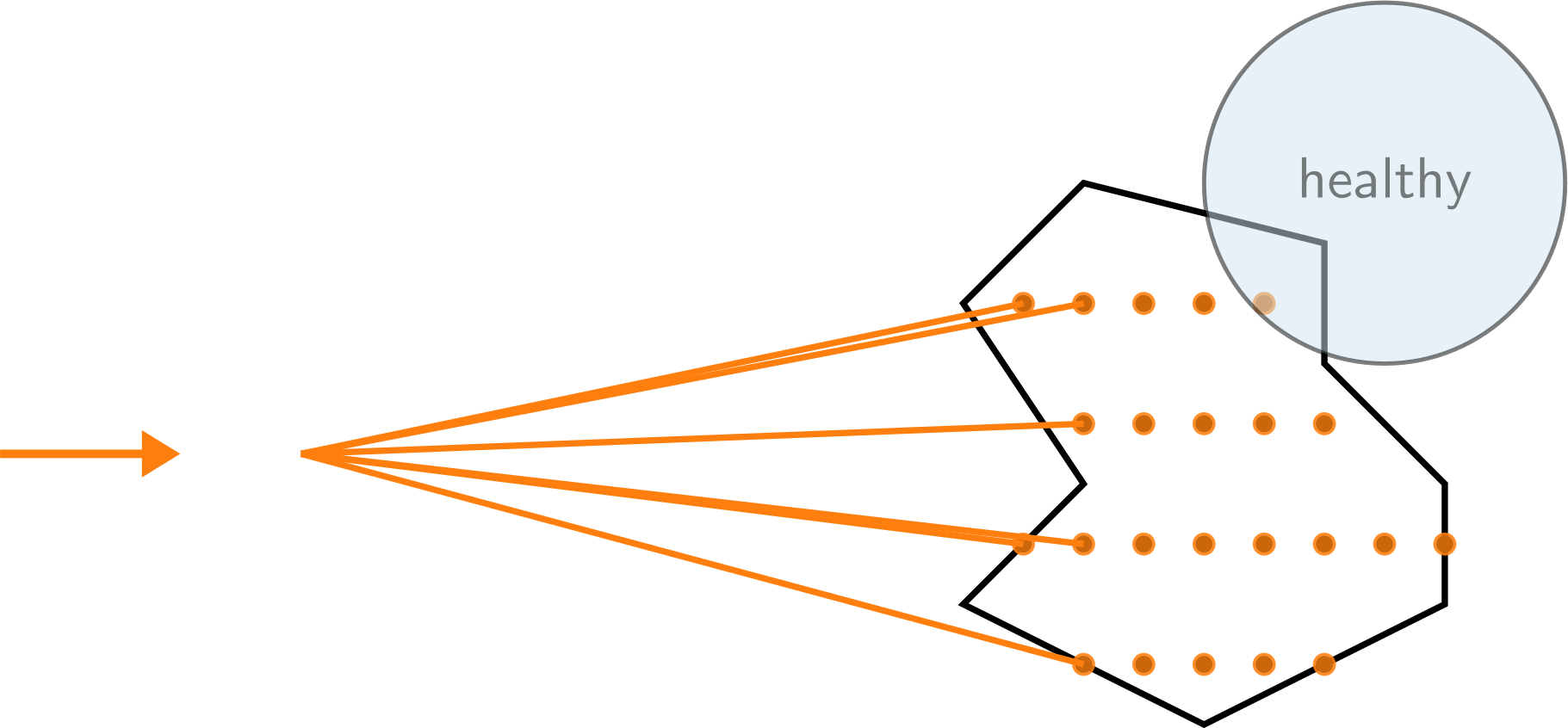}
    \caption{Schematic of pencil beam proton radiotherapy. The beam enters the patient from left to right and is subsequently scanned over the tumour (polygon) at discrete positions (orange dots) called spots.}
    \label{fig:pbs}
\end{figure}

So called Pencil Beam Scanning proton therapy works by scanning the tumor with a proton beam (see Fig.~\ref{fig:pbs}). The discrete positions at which the beam is turned on are called spots, and the on-times, called spot weights, need to be optimised to obtain a good dose distribution. In the example depicted in the figure, lower spot weights are preferable for spots within the healthy tissue that overlaps with the tumor. Spot weight optimisation algorithms tune the on-times for each spot using the influence matrix, which contains the dose contribution of each spot.  Finally, all of the decision making and validation of the created treatment plans is based on the dose. The JulianA.jl TPS interface consists of three features that need to be implemented for a given TPS: One for spot placement, one for calculating the so-called influence matrix and one for calculating the three-dimensional dose distribution within the patient.

All of these features need to be implemented in Julia for a given TPS to allow the creation and validation of a treatment plan. Concretely, each TPS is represented by a struct that is a subtype of \mintinline{julia}{Juliana.AbstractTps}. Such a struct configures the respective physics engine, e.\ g.\ select algorithm flavours, versions and other TPS-specific parameters. Three methods need to be specialised for a concrete type \mintinline{julia}{TpsType}
\begin{minted}{julia}
place_spots(
    tps::TpsType,
    ct,
    field_definitions,
    field_structures,
)
calculate_Dij_matrix(
    tps::TpsType,
    ct,
    plan,
    optimisation_points,
)
calculate_dose(
    tps::TpsType,
    resolution,
    ct,
    plan,
).
\end{minted}
Once these methods are implemented, the TPS's physics engines can be used transparently. All algorithms in JulianA.jl are built on top of these abstractions. Consequently, all scripts that use JulianA.jl can be adapted to another TPS without any modifications except to instantiate a different subtype of \mintinline{latex}{Juliana.AbstractTps}.

The TPS interface currently supports a command-line version of our in-house TPS FIonA and a default implementation written entirely in Julia, which can be used as a quick start without having to setup a research environment for a commercial TPS. We are investigating further TPS integrations for commercial systems. Collaborations for including other TPS systems are highly appreciated.

\section{Example code}
\begin{figure}
    \centering
    \includegraphics[width=0.475\textwidth]{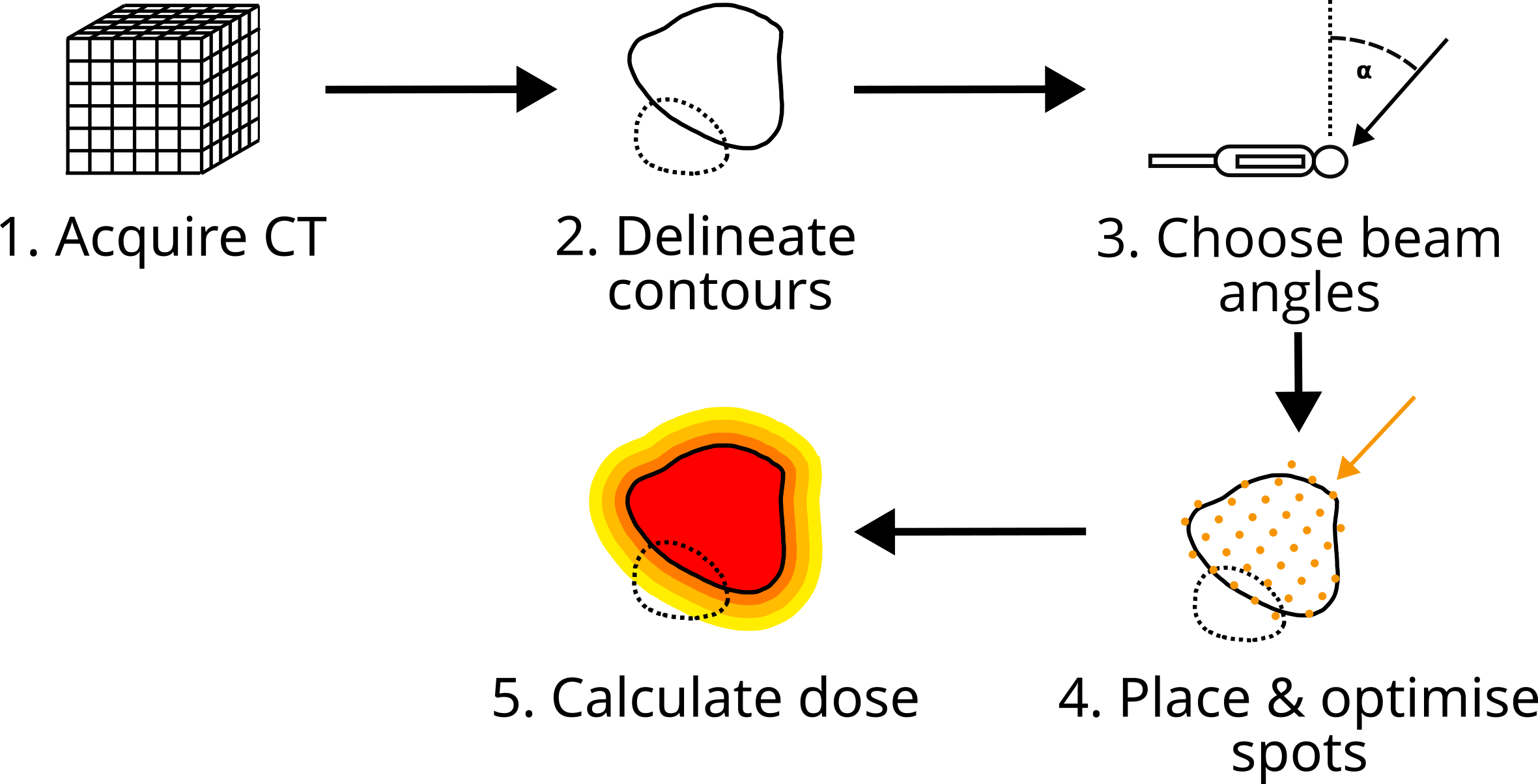}
    \caption{Schematic of the treatment workflow for radiotherapy.}
    \label{fig:workflow}
\end{figure}

Radiotherapy treatments employ the following workflow (see Fig.~\ref{fig:workflow}): Acquire CT, annotate the contours of the healthy organs and the tumours, define angles from which to irradiate (called fields) and the field centers, perform a spot weight optimisation, calculate the dose distribution and extract performance metrics. This workflow is implemented by JulianA.jl, but CT acquisition and contour delineation are replaced by loading the corresponding files from disk. The numbers in the comments for the following code correspond to the step numbers in Fig.~\ref{fig:workflow}:

\newpage
\begin{minted}[breaklines,escapeinside=||,mathescape=true, linenos, numbersep=3pt, gobble=2, frame=lines, fontsize=\small, framesep=2mm]{julia}
using Juliana
# Support for our in-house TPS.
using FionaStandalone

# 1. & 2. Load CT and contours.
ct, structures = Juliana.load_dicom_directory(data_dir)
constraints = Juliana.parse_oar_constraints_file(
    "constraint_file.csv",
    target_dose,
    structures,
)
prescriptions = Juliana.Prescriptions(
    [
        # my_tumor must receive 54.12Gy of dose.
        ("my_tumor", 54.12f0),
    ],
    constraints,
)
# 3. Choose beam angles.
fields, center_structures = load_fields(
    "beam_file.csv", structures,
)
# The only lines that are specific
# to the concrete TPS.
tps = FionaStandalone.FionaStandaloneTps(
    fiona_standalone_bin_path,
    fiona_jar_path,
    output_dir,
)
# 4. Place & optimise spots.
plan = Juliana.place_spots(
    tps,
    ct,
    fields,
    center_structures,
)
optimal_weights, _, _ = Juliana.optimise_head_and_neck(ct, structures, prescriptions, plan, tps, tps.work_dir)
plan_optimised = Juliana.update_spot_weights(
    plan,
    optimal_weights,
)
# 5. Calculate a dose distribution.
dose = Juliana.calculate_dose(
    tps,
    0.35f0, # dose grid resolution in cm
    ct,
    plan_optimised,
)
\end{minted}
The CT image and structure contours (tumor and healthy organs) are loaded from DICOM files, which are the standard in medical imaging. Next, the prescriptions are loaded and a connection to the TPS is established. In the case of our in-house system FIonA, this corresponds to running a command line executable, but other TPS implementations might rely on REST calls or other approaches. Afterwards, the spot placement and the spot weight optimisation are performed to obtain a treatment plan. The final steps are to calculate the full dose distribution and calculate the performance metrics for clinical plan approval. All of these steps are fully automatic thanks to the included autoplanning algorithm \cite{bellotti2023} and are agnostic of the concrete TPS. The TPS interface needs to be implemented once, but afterwards the algorithms that rely on it are portable across vendors and even institutes. TPS integration and algorithms can be developed in parallel and are orthogonal to each other, which boosts productivity.

Due to space limitations, we are limited to displaying only the small code snippet above. However, JulianA.jl also contain a vast array of utility functions for calculating distance-to-structure, binary masks, boolean operations on structures, calculation of beam entry regions and radiological path length, plotting of dose distributions and performance metrics and even generating full \LaTeX reports for a given treatment plan. All of the functionality is open source \cite{code}.

\begin{figure}
    \centering
    \includegraphics[width=0.45\textwidth]{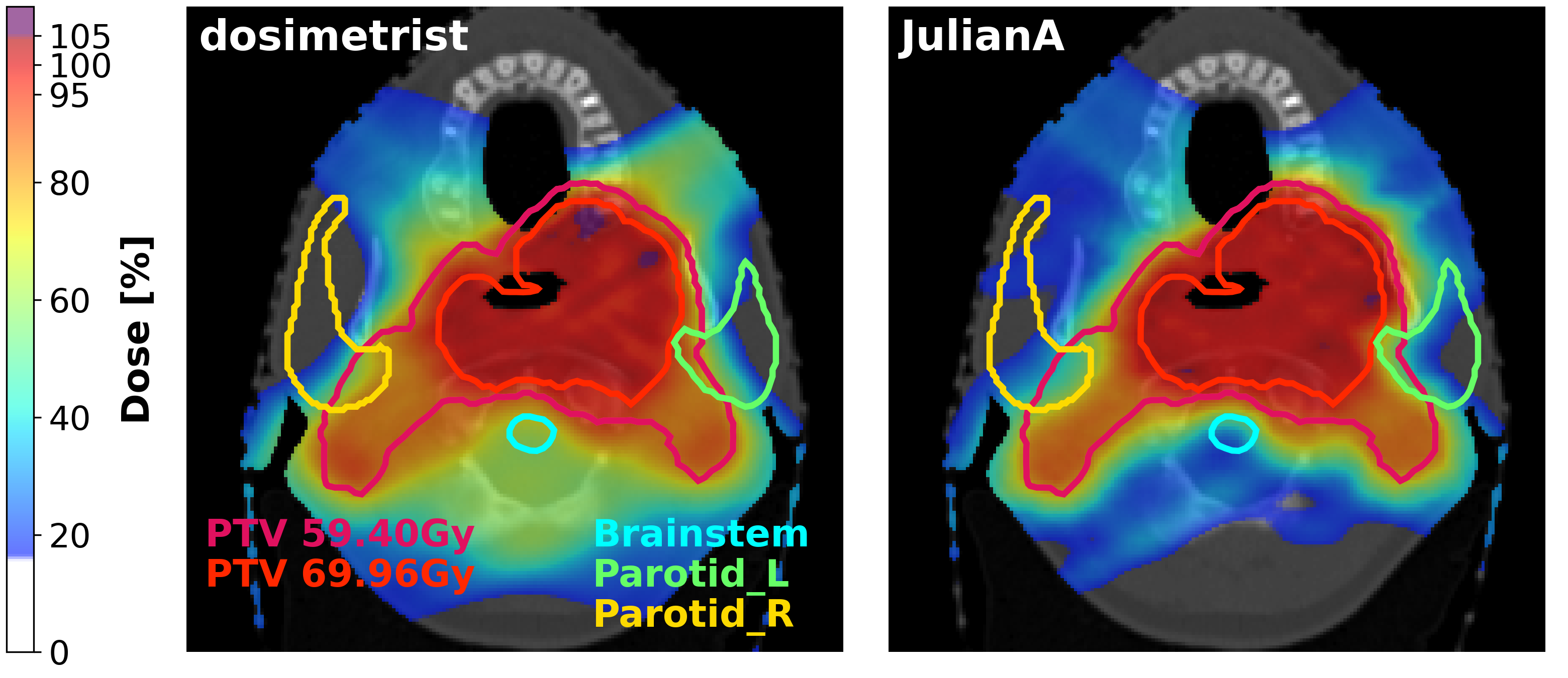}
    \caption{Dose distribution created by a dosimetrist and accepted for treatment (left) and dose distribution generated fully automatically using JulianA.jl (right) for a nasopharyngeal cancer patient treated at our institute. The contours represent the tumors (called planning target volume PTV) subdivided by prescribed dose in Gy, and healthy organs (delineated in green, yellow and cyan) close to the target. Color wash isodose levels of the prescribed dose are displayed on the left-hand side of the figure.}
    \label{fig:example_dose}
\end{figure}
A case study is presented in Fig.~\ref{fig:example_dose}. The plot on the left shows the dose distribution obtained by a human treatment planner the plot on the right shows the solution found automatically by JulianA.jl in approx.~\SI{10}{min}. Both treatment plans are of a comparable quality, but JulianA.jl achieves a better sparing of the normal tissue close to the brainstem and the back of the head, which reduces potentially the risk of radiation-induced side effects. A validation study involving two radiation oncologists is under preparation. The preliminary results show that JulianA.jl generates plans acceptable for treatment for $17$ (out of $17$) patients. For $13$ cases, the JulianA.jl plan is considered superior to the human-made treatment plan, for $1$ case both plans are considered to be of equal quality and for $2$ cases, the human planners beat JulianA.jl.

\section{Applications}
The JulianA.jl package has already proven useful within our institute. Thanks to the versatility of the framework and the simplicity of the Julia language, implementing the novel autoplanning algorithm for spot weight optimisation (now included in JulianA.jl) was straightforward and allowed us to focus on tuning the algorithm rather than dealing with technicalities. For example, we could rely on a vast selection of optimisation algorithms from the \texttt{Optim.jl} package \cite{optim}. Further, we have applied JulianA.jl to create a fully automatic pipeline for the validation of a machine learning model \cite{li2024}. The development of the pipeline took only a single day and it could be run fully automatically. Without JulianA.jl, each iteration in the development of the model would have needed several days of validation and manual treatment planning. Finally, we are currently exploring the utility of JulianA.jl for QA in our daily clinical workflows and for automatic planning of head-and-neck cancer patients.

% \begin{itemize}
%     \item Explain workflow within a TPS and highlight how the code example follows this workflow 1:1
%     \item Example code for loading, calculating binary masks, optimising, calculating a dose distribution, plotting and reporting
%     \item Show example treatment plan
%     \item Highlight that no manual interaction is needed and the resulting DICOM are readable by any TPS and other medical software
%     \item TPS interface needs to be developed once, but this is orthogonal to the development of algorithms, which allows separate devlopment and follows the single-responsibility principle
% \end{itemize}

\section{Conclusion and Outlook}
We have developed JulianA.jl as a simple and efficient Julia package for radiotherapy. It unlocks the power of modern software and hardware stacks for the radiotherapy community and aims to boost productivity and efficiency in research by fostering code reuse within and across institutes, but also provide vendor-agnostic scripting capabilities for the clinics.

We are currently writing a validation study to explore the utility of our autoplanning algorithm for clinical applications for head-and-neck cancer patients. Further, we are planning to use JulianA.jl to accelerate multiple research endeavors at our institute, especially in the realm of machine learning. Further, we are integrating JulianA.jl with a commercial TPS, extending the documentation and writing automatic tests.

\section{Acknowledgements}
This work has been done as part of the INSPIRE project and is supported by a PSI CROSS project.

We acknowledge the assistance of Derek Feichtinger and Marc Caubet for their help with the Merlin and Gwendolen cluster, which enabled the computational work of the research.

\def\refname{References}

\vspace*{-8pt}

\end{document}